\documentclass[utf8]{FrontiersinHarvard} 

\usepackage{url,hyperref,lineno,microtype,subcaption}
\usepackage[onehalfspacing]{setspace}

\def\keyFont{\fontsize{8}{11}\helveticabold }
\def\firstAuthorLast{Sample {et~al.}} 
\def\Authors{Talgar Bayan\,$^{1}$, Richard Banach\,$^{1}$, Askar Nurbekov \,$^{2}$, Makhmud Mustafabek Galy$^{4}$, Adi Sabyrbayev $^{5}$, Zhanat Nurbekova\,$^{3,*}$}
\def\Address{
$^{1}$Department of Computer Science, Manchester, UK \\
$^{2}$Department of Computer Science, The University of Texas Rio Grande Valley, Edinburg, US \\
$^{3}$ Department of Informatics and Informatization of Education, Abai Kazakh National Pedagogical University, Almaty, KZ \\
$^{4}$ Data Platform Department, Opn Co. Ltd., Bangkok, TH \\
$^{5}$ Google, Warsaw, PL
}
\def\corrAuthor{Zhanat Nurbekova}

\def\corrEmail{nurbekova\_zk@digitalexgroup.com}

\begin{document}
\onecolumn
\firstpage{1}

\title[Running Title]{Blockchain-enhanced Integrity Verification in Educational Content Assessment Platform: A Lightweight and Cost-Efficient Approach} 


\author[\firstAuthorLast ]{\Authors} 
\address{} 
\correspondance{} 

\extraAuth{}

\maketitle

\begin{abstract}
The growing digitization of education presents significant challenges in maintaining the integrity and trustworthiness of educational content. Traditional systems often fail to ensure data authenticity and prevent unauthorized alterations, particularly in the evaluation of teachers' professional activities, where demand for transparent and secure assessment mechanisms is increasing. In this context, Blockchain technology offers a novel solution to address these issues. This paper introduces a Blockchain-enhanced framework for the Electronic Platform for Expertise of Content (EPEC), a platform used for reviewing and assessing educational materials. Our approach integrates the Polygon network, a Layer-2 solution for Ethereum, to securely store and retrieve encrypted reviews, ensuring both privacy and accountability. By leveraging Python, Flask, and Web3.py, we interact with a Solidity-based smart contract to securely link each review to a unique identifier (UID) that connects on-chain data with real-world databases. The system, containerized using Docker, facilitates easy deployment and integration through API endpoints. Our implementation demonstrates significant cost savings, with a 98\% reduction in gas fees compared to Ethereum, making it a scalable and cost-effective solution. This research contributes to the ongoing effort to implement Blockchain in educational content verification, offering a practical and secure framework that enhances trust and transparency in the digital education landscape.

\tiny
 \keyFont{ \section{Keywords:} Blockchain, Smart Contract, Content Verification, Layer 2, Polygon Network, Data Encryption, Educational Content, Gas Optimization} 
\end{abstract}

\section{Introduction}
The digital transformation of education has introduced new challenges in ensuring the integrity and quality of educational content \cite{oliveira2022digital,chatgpt2023,sefcik2019mapping}. As institutions move towards Education 4.0, the need for robust mechanisms to verify and maintain the integrity of digital educational materials has become increasingly critical \cite{dawson2020defending,khan2023connecting,roe2022automated}. Traditional methods of content evaluation often lack robust mechanisms to prevent unauthorized alterations, particularly in the review of critical educational materials \cite{khan2023connecting,roe2022automated}. By analyzing existing research, we have identified key knowledge gaps and opportunities for strengthening the digital ecosystem and supporting continuous professional development in education. 

\begin{figure}[h!]
\begin{center}
\includegraphics[width=16cm]{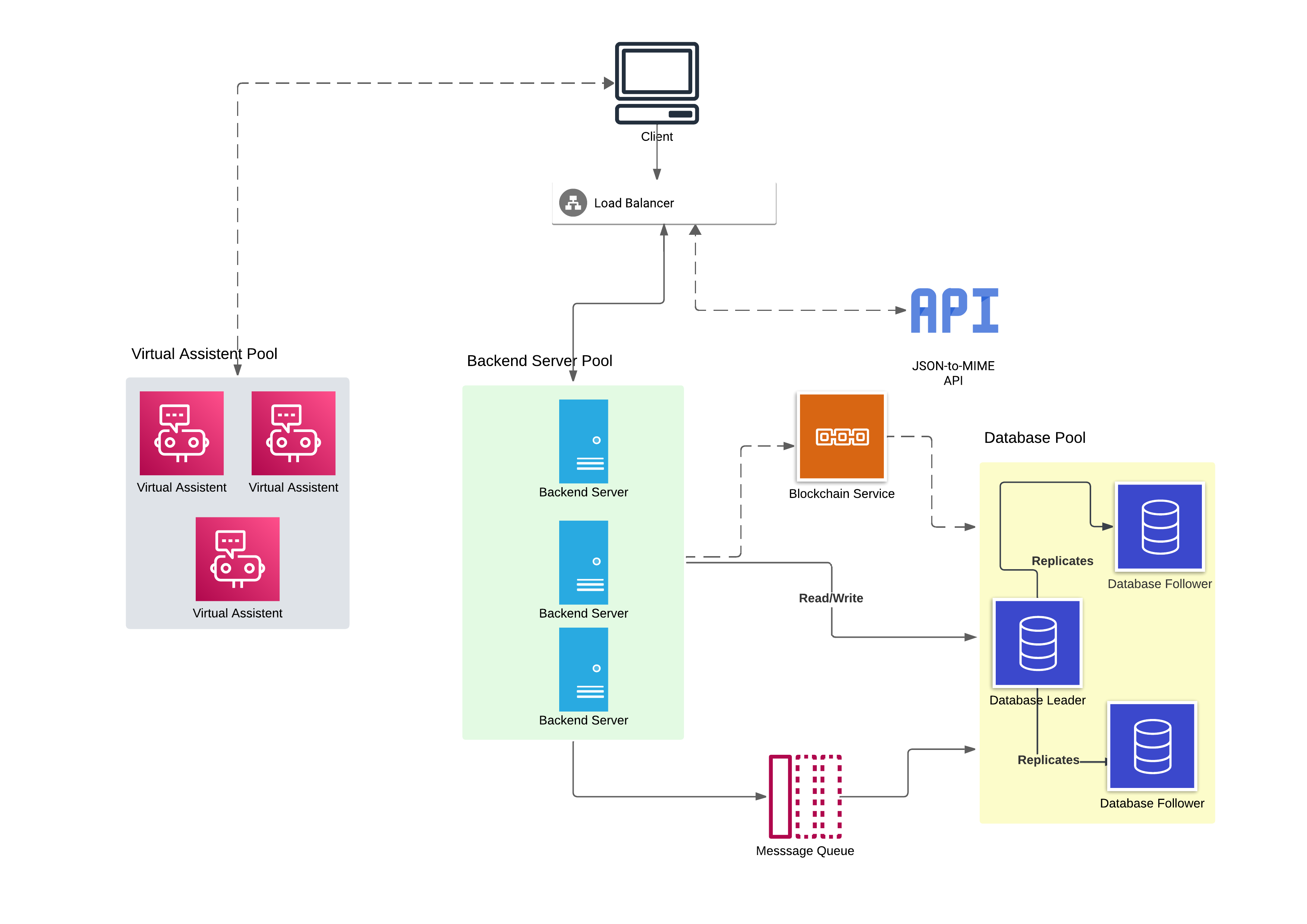}
\end{center}
\caption{High-Level Architecture of EPEC Platform.}
\label{fig:flowchart}
\end{figure}

The Electronic Platform for Expertise of Content (EPEC) is a multi-layered software solution designed to automate the process of reviewing and assessing the quality of educational materials. Developed as part of the scientific program "Scientific Foundations for the Modernization of Education and Science," EPEC offers a multi-layered software approach to provide potential solutions to the above-mentioned challenges. Recent advancements in the EPEC project have further enhanced its capabilities. Nurbekova, Sembayev et al. \cite{nurbekova2023multi} introduced a multi-criteria-based expert modelling system for textbook quality assessment, demonstrating the potential for increased rigour and transparency in content evaluation. Building on this, Nurbekova, Aimicheva et al. \cite{nurbekova2023decision} presented a decision-making platform for educational content assessment within a stakeholder-driven digital ecosystem, showcasing the practical application of advanced technologies in educational assessment systems. EPEC is now fully functional and operational. Technically, EPEC uses a lightweight and modern tech stack: a RESTful API, Django Rest Framework, Postgres 15 DBMS, MinIO Object, JSON-to-MIME, RabbitMQ, and a Virtual Assistant built using modern Large Language Models (LLMs) trained on specific datasets. Figure \ref{fig:flowchart} illustrates the architecture of EPEC. Notably, the diagram highlights the modular, detachable Blockchain service, which is distinctly labeled in orange.

Recent studies have highlighted the potential of Blockchain technology in addressing challenges in educational settings \cite{steiu2020Blockchain,bhaskar2020Blockchain,ocheja2022Blockchain}. Blockchain's inherent characteristics of immutability, transparency, and decentralization make it a promising solution for enhancing the integrity and trustworthiness of educational content \cite{ramos2020Blockchain,loukil2021Blockchain}. Several researchers have explored the applications of Blockchain in education management, including areas such as credential verification, record-keeping, and content authentication \cite{bhaskar2020Blockchain,ocheja2022Blockchain,tariq2023cerberus}.

This paper presents a novel approach to integrating Blockchain technology with EPEC, specifically focusing on the use of the Polygon network, a Layer 2 scaling solution for Ethereum. Our work makes several key contributions to the field of Blockchain applications in educational technology:
\begin{enumerate}
    \item We provide a comparison of Blockchain platforms, identifying a low-cost, lightweight solution suitable for educational content management. This analysis leads to the selection of an optimal tech stack that ensures safe and effective implementation, addressing the critical need for cost-effective Blockchain solutions in education \cite{bayan2024privacypreserving,zarir2021developing}.
    \item We develop a flexible integration architecture using Docker containerization, Flask for web application development, and Web3.py for Blockchain interaction. This design allows for modular integration of Blockchain functionality with existing systems like EPEC, enabling easy activation or deactivation of Blockchain features as needed. This approach significantly lowers the barrier to Blockchain adoption in educational technologies \cite{jameel2024transforming}.
    \item We implement and evaluate a smart contract system for storing and retrieving encrypted messages, demonstrating a practical approach to maintaining data privacy while leveraging Blockchain's immutability. This implementation, interfacing seamlessly with a Flask application through Python, provides a replicable model for secure, Blockchain-based data management in educational contexts \cite{ertan2019using,khan2021Blockchain}.
\end{enumerate}

\section{Related Works}
\subsection{Blockchain Technology in Education}
The application of Blockchain technology in education has garnered significant attention from researchers and practitioners alike. Grech and Camilleri \cite{grech2017Blockchain} provided a comprehensive overview of Blockchain's potential to transform administrative processes, certification, and overall transparency in academic activities. Their research highlighted Blockchain's capacity to enhance data security, ensure the immutability of records, and create a reliable system for credential verification. Sharples and Domingue \cite{sharples2016Blockchain} explored the transformative potential of Blockchain for lifelong learning. They presented a framework allowing clients to store their achievements, credentials, and learning records on the Blockchain. This study demonstrated how decentralized, permanent records could ensure the integrity of educational achievements and assessments across various levels of education.

Recent studies have further expanded our understanding of Blockchain applications in education. For instance, Tariq et al. \cite{tariq2023cerberus} proposed Cerberus, a Blockchain-based system for accreditation and degree verification. Their work demonstrates the practical implementation of Blockchain technology in educational credentialing, addressing issues of forgery and verification inefficiencies. Similarly, Rustemi et al. \cite{rustemi2023systematic} conducted a systematic literature review on Blockchain-based systems for academic certificate verification, providing a comprehensive overview of current approaches and challenges in this domain.

\subsection{Smart Contracts in Educational Assessment}
Smart contracts, self-executing contracts with the terms of the agreement directly written into code, have been recognized as a key tool in automating educational assessments. Zheng et al. \cite{zheng2020overview} explored the potential of smart contracts in educational institutions, noting that they could streamline processes such as certification and evaluation. In the context of educational assessment, smart contracts can automate the evaluation process, automatically triggering reviews after certain conditions are met. Khan et al. \cite{khan2021Blockchain} provided a comprehensive review of Blockchain smart contract applications, challenges, and future trends, including their potential in educational contexts. Ren et al. \cite{ren2021empirical} conducted an empirical evaluation of smart contract testing methods, providing insights into best practices for ensuring the reliability and security of smart contracts in critical applications like educational assessment.

\subsection{Blockchain-based Content Verification}
The potential of Blockchain to create transparent and decentralized assessment processes was noted by Alammary et al. \cite{alammary2019Blockchain}. They argued that Blockchain could provide secure, decentralized platforms for storing and managing assessment data, potentially helping to eliminate bias and standardize assessment methods across institutions. This concept is particularly relevant for educational content verification, where subjective evaluation criteria and inconsistent assessment methods often compromise the accuracy and fairness of content reviews \cite{hershkovitz2020role}. Erdayandi et al. \cite{erdayandi2023privacy} proposed a privacy-preserving and accountable billing protocol for peer-to-peer energy trading markets, utilizing Homomorphic encryption and Blockchain technology to ensure data security and transaction integrity. 

While not specifically focused on Blockchain, Song et al. \cite{song2022application} demonstrated the application of web-based hazard maps in high school education, highlighting the importance of reliable and verifiable digital content in educational settings. This study underscores the potential for Blockchain-based solutions to enhance the integrity and reliability of educational content across various disciplines.

As Blockchain adoption in education grows, Layer 2 solutions, such as the Polygon network, offer promising avenues for overcoming these challenges. However, research on the specific application of Layer 2 solutions in educational contexts remains limited, representing a significant gap in the literature that our study aims to address. Masla et al. \cite{masla2021reduction} and Zarir et al. \cite{zarir2021developing} have explored techniques for reducing gas costs in Blockchain-enabled smart contracts. Their work provides valuable insights into optimizing the economic efficiency of Blockchain applications, which is particularly relevant for educational institutions operating under budget constraints. Marchesi et al. \cite{marchesi2020design} proposed design patterns for gas optimization in Ethereum, offering practical strategies for developing cost-effective Blockchain applications. While their work is not specific to educational applications, the principles they outline are highly relevant to our aim of creating a lightweight and cost-efficient Blockchain solution for educational content assessment.

\subsection{Research Gap and Our Contribution}
Our work aims to address this gap by providing a comprehensive implementation and evaluation of a Blockchain-based integrity verification system for educational content using the Polygon network.
Furthermore, our work extends beyond theoretical propositions by providing a practical implementation and evaluation of smart contracts for educational content verification. This hands-on approach contributes valuable insights into the real-world applicability and challenges of Blockchain technology in educational settings, paving the way for future research and development in this critical area.

\section{Methodology and Implementation Details}
Our approach to implementing Blockchain-enhanced integrity verification in EPEC involves several key steps, from the selection of an appropriate Blockchain platform to the design and implementation of smart contracts and their integration with EPEC. This section outlines our methodological approach and provides details on its implementation.
We use a \textbf{microservice architecture} and modular coding for better scaling, maintenance and implementation purposes. These involve docker technology, APIs, back end and front end, and the use of web3 packages to interact with the on-chain data. Our tasks include:
\begin{itemize}
    \item Develop a secure method for storing and retrieving encrypted reviews on the Polygon Blockchain.
    \item Implement the solution using Python, Flask, and Web3.py for ease of integration with web applications.
    \item Associate each review with a UID to link on-chain data with real-world data.
    \item Containerize the application using Docker for easy deployment and integration with external systems through API endpoints.
\end{itemize}

\subsection{Blockchain Platform Selection and Rationale}
The Blockchain ecosystem has grown significantly, encompassing both permissioned blockchains like Hyperledger Fabric and Corda, and permissionless blockchains such as Ethereum, Polygon, Solana, and Arbitrum. With thousands of developers and millions of users interacting with these chains, they have been thoroughly tested and verified. These chains are suitable for industrial-level applications and services, offering features like immutability and decentralization. The computational and financial costs required to compromise such a robust Blockchain system are exponentially higher than the value of the data itself, effectively deterring malicious attempts to hack or alter the stored information.

Our previous work provides a detailed comparison of different Blockchain platforms, highlighting key differences in properties such as security level, transaction fees, and scalability. Table \ref{table_comparison} presents a comparative analysis of various Blockchain platforms, including Ethereum, Polygon, Hyperledger Fabric, and Private Custom chains \cite{bayan2024privacypreserving}.

\begin{table*}[h!]
\renewcommand{\arraystretch}{1.3} 
\caption{Comparative Analysis of Blockchain Technologies (adapted from \cite{bayan2024privacypreserving})}
\label{table_comparison}
\centering
\begin{tabular}{|m{0.20\linewidth}|m{0.15\linewidth}|m{0.15\linewidth}|m{0.15\linewidth}|m{0.15\linewidth}|}
\hline
\textbf{Properties} & \textbf{Ethereum} & \textbf{Polygon} & \textbf{Hyperledger Fabric} & \textbf{Private Custom Chain} \\ \hline
Type & Permissionless, Decentralized & Permissionless, Decentralized & Permissioned, Centralized & Permissioned, Centralized \\ \hline
Security Level & High & High & Moderate & Moderate \\ \hline
Transaction Fee & High & Low & Low & Low \\ \hline
Operational Expenditure (Server, Maintenance, Energy, Development) & Low & Low & High & Very High \\ \hline
Ease of Implementation & Moderate & Moderate & High & Very High \\ \hline
Large Institutions Involvement & Yes & Yes & Yes & Yes (Primarily for CBDCs) \\ \hline
Transaction Speed & Moderate & High & Very High & Very High \\ \hline
Regulatory Status & In Process & In Process & Regulated & Regulated \\ \hline
Scalability & Moderate & High & Moderate & Low \\ \hline
Consensus Mechanism & Proof of Work (Transitioning to Proof of Stake) & Proof of Stake & Pluggable (e.g., PBFT) & Varies (Often Custom) \\ \hline
Developer Community & Extensive & Growing & Moderate & Limited \\ \hline
\end{tabular}
\end{table*}

Ethereum is the most advanced and widely developed blockchain, but it faces issues like scalability and high gas fees. To tackle these challenges, Layer 2 blockchains have been developed to handle tasks offloaded from Ethereum, enabling faster and cheaper transactions. Popular Layer 2 solutions such as Arbitrum, Polygon, and Optimism benefit from being secured by the Ethereum network, making them reliable platforms for further development. We chose Polygon as the foundation for our service for several reasons. First, Polygon offers \textbf{low transaction costs}, with fees starting as low as USD 0.015 per transaction, making it a cost-effective solution for recording book reviews on the blockchain. Additionally, \textbf{Polygon has been widely adopted} by large institutions and even governments, which shows its reliability and scalability. Polygon's \textbf{compatibility with the Ethereum Virtual Machine (EVM)} ensures that our service can easily integrate into the Ethereum ecosystem. The \textbf{high transaction speed} offered by Polygon is essential for efficient data management, and its \textbf{scalability} ensures that our service can grow without compromising on security. Moreover, Polygon's \textbf{eco-friendly approach} fits well with our sustainability goals, and its \textbf{developer support} provides extensive resources for building and maintaining our service.

Our service is designed to be lightweight and cost-effective. For example, storing the final reviews for 2000 books will cost less than USD 100, significantly reducing ongoing costs. To add flexibility, our service includes a \textbf{'global switch'} feature, which allows the blockchain component to be attached or detached from the main system as needed, without impacting security or functionality. The service is also Dockerized, making it easier to maintain and connect, ensuring a more adaptable and user-friendly solution.

\subsection{Smart Contract Design and Implementation}
\begin{figure}[h!]
\begin{center}
\includegraphics[width=16cm]{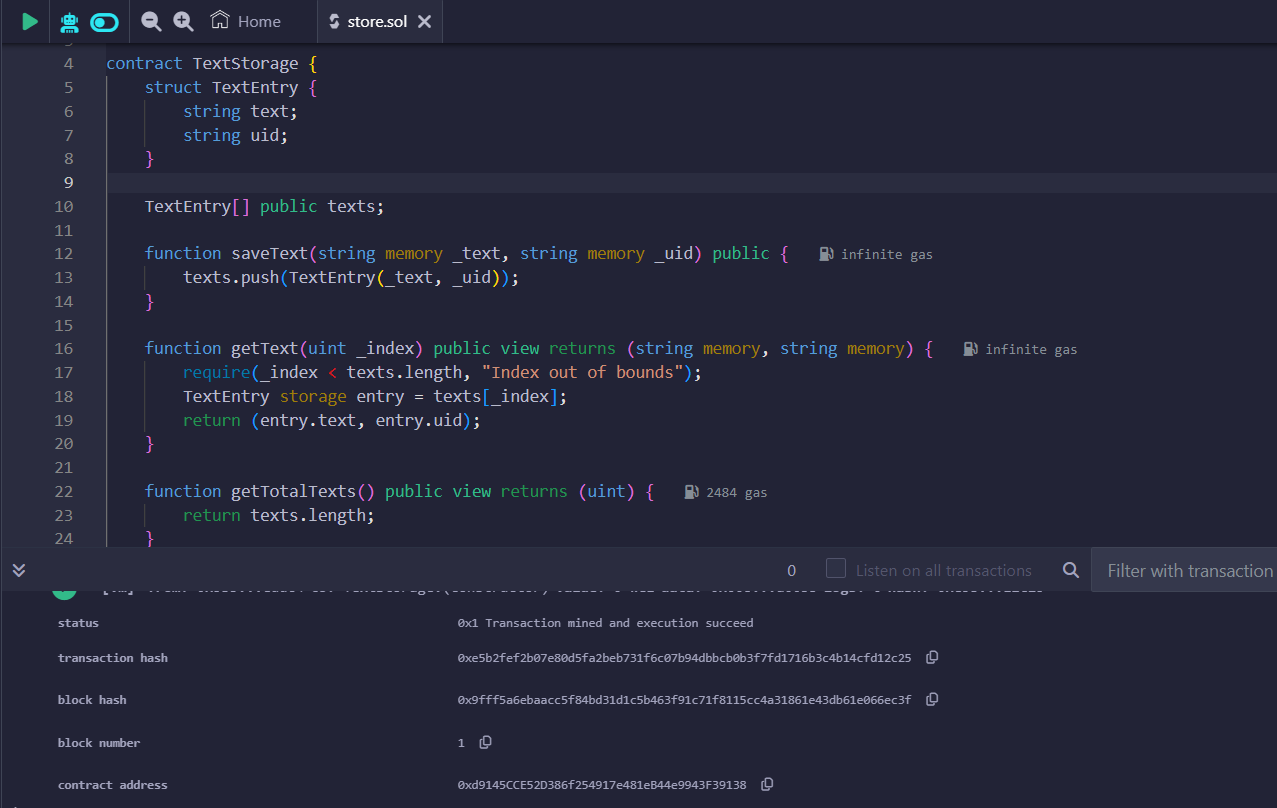}
\end{center}
\caption{ Functionality of our TextStorage smart contract.}\label{fig:smart_contract_textStorage}
\end{figure}

Our Blockchain implementation utilizes a smart contract written in Solidity, a high-level language specifically designed for Ethereum-based platforms. The smart contract, named TextStorage, is intentionally designed with simplicity and security in mind to mitigate potential audit issues and security vulnerabilities.
The contract's structure is based on a single struct called TextEntry, which encapsulates two string variables: text and uid. This design allows for the storage of encrypted text alongside a unique identifier, facilitating efficient retrieval and verification processes. The contract implements three primary functions. The saveText function accepts two parameters, \_text and \_uid, and appends a new TextEntry to the texts array. This operation allows for the sequential storage of encrypted text entries on the Blockchain. The getText function, given an index, retrieves the corresponding TextEntry from the texts array. It includes a boundary check to prevent out-of-range access attempts, enhancing the contract's robustness. Lastly, the getTotalTexts function returns the total number of stored text entries, providing a simple mechanism for tracking the contract's state. Figure \ref{fig:smart_contract_textStorage} presents the core functionality of our TextStorage smart contract. 

By leveraging the immutable nature of Blockchain technology, our smart contract ensures that once data is recorded, it cannot be altered or tampered with. This immutability, combined with the transparency of all transactions on the Blockchain, provides a high degree of security and trust, making it an ideal platform for storing critical, encrypted information. The contract's simplicity serves multiple purposes. It minimizes the attack surface, reducing potential vulnerabilities. It also optimizes gas consumption, a critical factor in Blockchain operations. Furthermore, it facilitates easier auditing and verification processes, crucial for maintaining the integrity of the system. In our implementation, we store the contract's Application Binary Interface (ABI) locally within our development environment. This approach enables seamless integration with Python and Web3.py, facilitating efficient interaction with the deployed smart contract on the Blockchain.

\subsection{Integration with EPEC}
\begin{figure}[h!]
\begin{center}
\includegraphics[width=12cm]{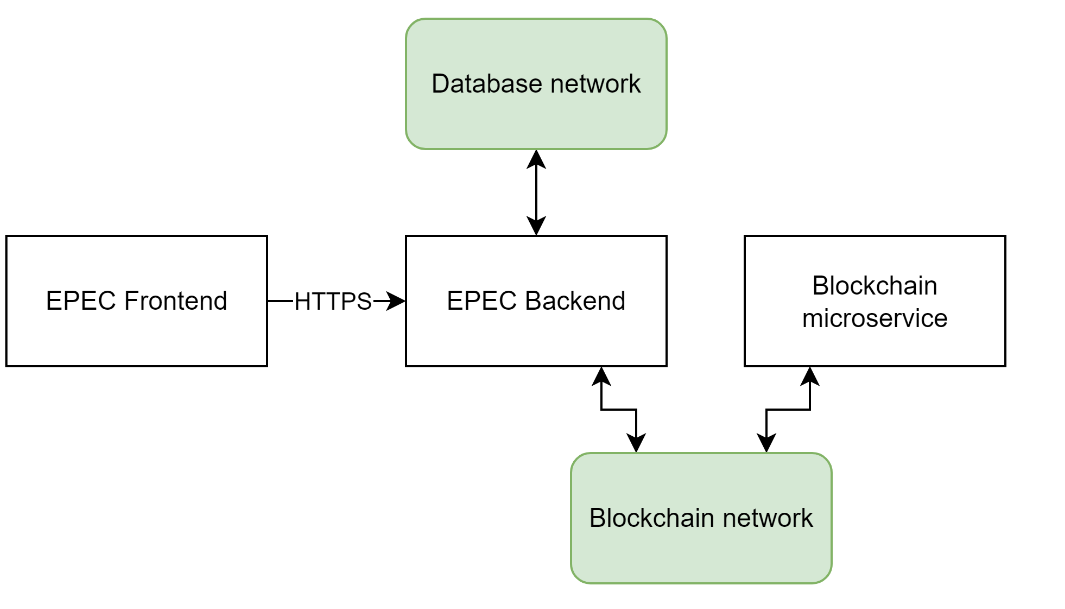}
\end{center}
\caption{ Database relations.}\label{fig:docker}
\end{figure}
\begin{figure}[h!]
\centering
\includegraphics[width=\textwidth]{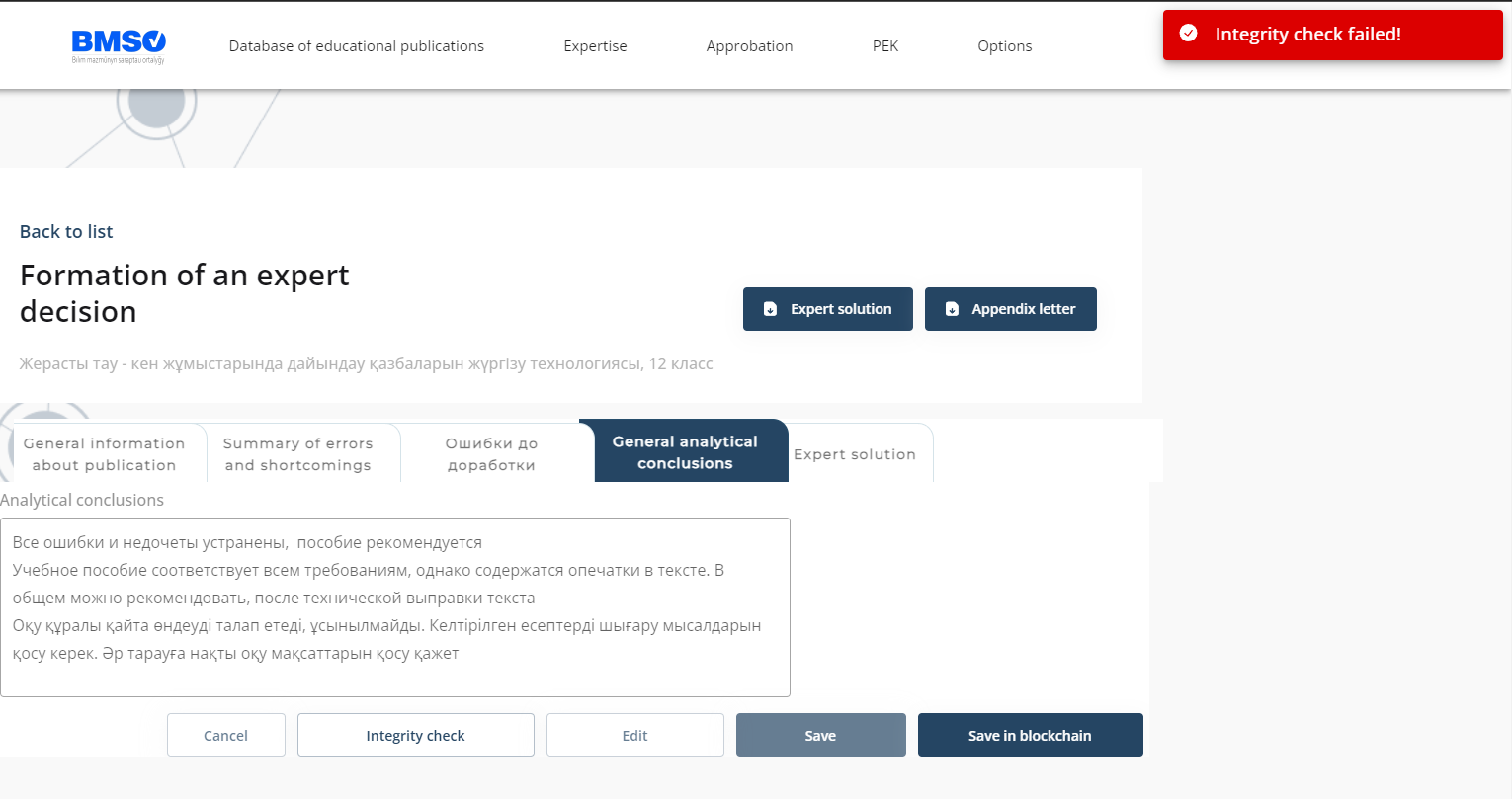}
\caption{User interface displaying a failed integrity check, indicated by a red notification when local data does not match the Blockchain record.}
\label{fig:integrity_check_failed}
\end{figure}

EPEC's primary system operates with a PostgreSQL database to manage local data and operations. By integrating this system with the on-chain smart contract, we enhance data integrity and security. Specifically, saving critical data such as examiners' reviews on both local and on-chain platforms enables us to achieve encrypted data storage. This dual storage approach provides the capability to verify whether local data has been tampered with. Local data, due to its centralized management, is susceptible to alteration. In contrast, on-chain data is immutable, making it resistant to unauthorized changes. By cross-verifying data between the two sources, we can easily detect discrepancies between the original and potentially altered data. The database inside the Docker networks is described below in Figure \ref{fig:docker}.
In the EPEC system, the integration of Blockchain services introduces a new functionality termed "Check for Integrity." This feature allows users to compare data stored locally with its original version on the Blockchain. When the comparison indicates data consistency, a green notification is displayed, as illustrated in Figure \ref{fig:green_noti} below.
\begin{figure}[h!]
\begin{center}
\includegraphics[width=16cm]{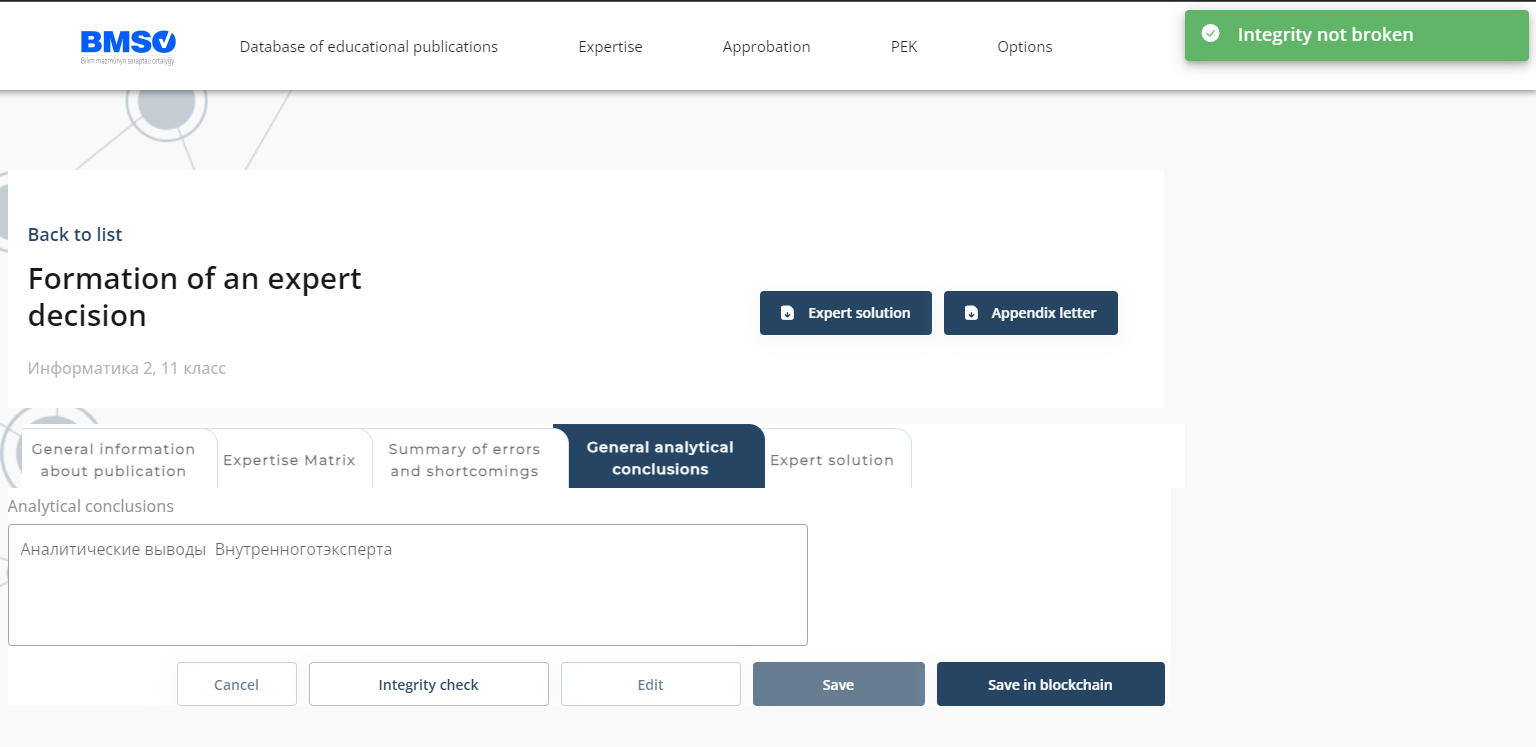}
\end{center}
\caption{ Integrity checking with the on-chain encrypted data and returned successful status.}\label{fig:green_noti}
\end{figure}
In cases where the integrity check detects a discrepancy between the local data and its on-chain counterpart, the system alerts users to potential data alteration. Figure \ref{fig:integrity_check_failed} illustrates the interface response when the integrity verification fails, specifically for expert reviews of educational content. The system compares local data with the Blockchain version using the associated unique identifier (UID) for each review. Upon detecting a mismatch, it displays a red notification, signaling a failed integrity check to users and alerting relevant staff members.
This integrity verification process can be automated through scheduled cron tasks, providing continuous monitoring of sensitive data and safeguarding against unauthorized alterations. Such a mechanism enhances the overall security and reliability of the EPEC system, ensuring the authenticity of expert reviews and other critical educational content assessments.

We have included the following measures to create and deploy a highly secure and audited smart contract:
\textbf{Access Control: }The smart contract utilizes the only owner modifier to restrict certain functions, such as saveText and transferOwnership, to the address that deployed the contract (owner). This access control mechanism ensures that only authorized parties can modify critical aspects of the contract, reducing the risk of unauthorized changes or malicious activities.
\textbf{Event Logging:} The contract emits events (TextAdded and OwnershipTransferred) to log important state changes and actions performed on the contract. Event logging enhances transparency and provides a means for external parties to monitor the contract's behaviour, facilitating auditing and debugging processes.
\textbf{No extra calls: } The provided contract does not interact with external contracts or external data sources (e.g., oracles) within its critical functions. Minimizing reliance on external calls can reduce the attack surface and potential vulnerabilities associated with interacting with untrusted oracles or contracts.

Using the latest version of solidity, and using minimalism and robust structure, to make sure we have a secure Smart Contract can be used. The main projects use the docker technology, and each service connects using an API via an endpoint, to achieve flexibility and ease of maintenance.
We use a Flask framework to write our backend parts, inside the Flask, we use Python and web3 packages to connect to the Blockchain. The flask app will be encapsulated inside a docker. all the variables will be stored inside a .env file. 

We use the post method to save the text from the user input, combine it with the unique UID (user identifier), generate a transaction, sign the transaction and then send it to the chain for broadcasting.  at the same time, catching any possible errors. During this process, the UID and generated transaction hash will be also saved in local databases, it will displayed when the user reads the reviews, easy for the user to retrieve themself if needed.  Also, they can retrieve and check the integrity with one click of the button we prepared. The user also has the ability to retrieve a transaction. It returns the status of a transaction that is posted to the Blockchain.

\begin{figure}[h!]
\begin{center}
\includegraphics[width=16cm]{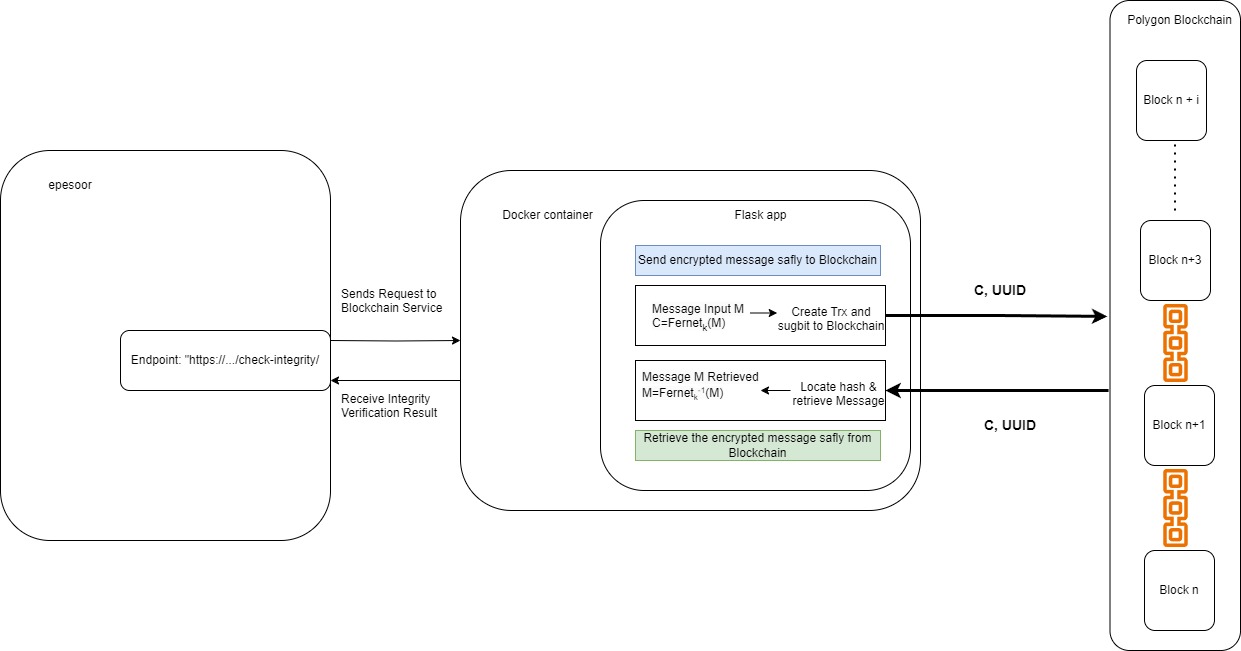}
\end{center}
\caption{ Flowchart of EPEC and Blockchain Service.}\label{fig:flowchart_connect}
\end{figure}

All the text will use Python's fernet encryption and description methods. And the minimized process flow can be found in this Figure\ref{fig:flowchart_connect}.

\begin{figure}[h!]
\begin{center}
\includegraphics[width=16cm]{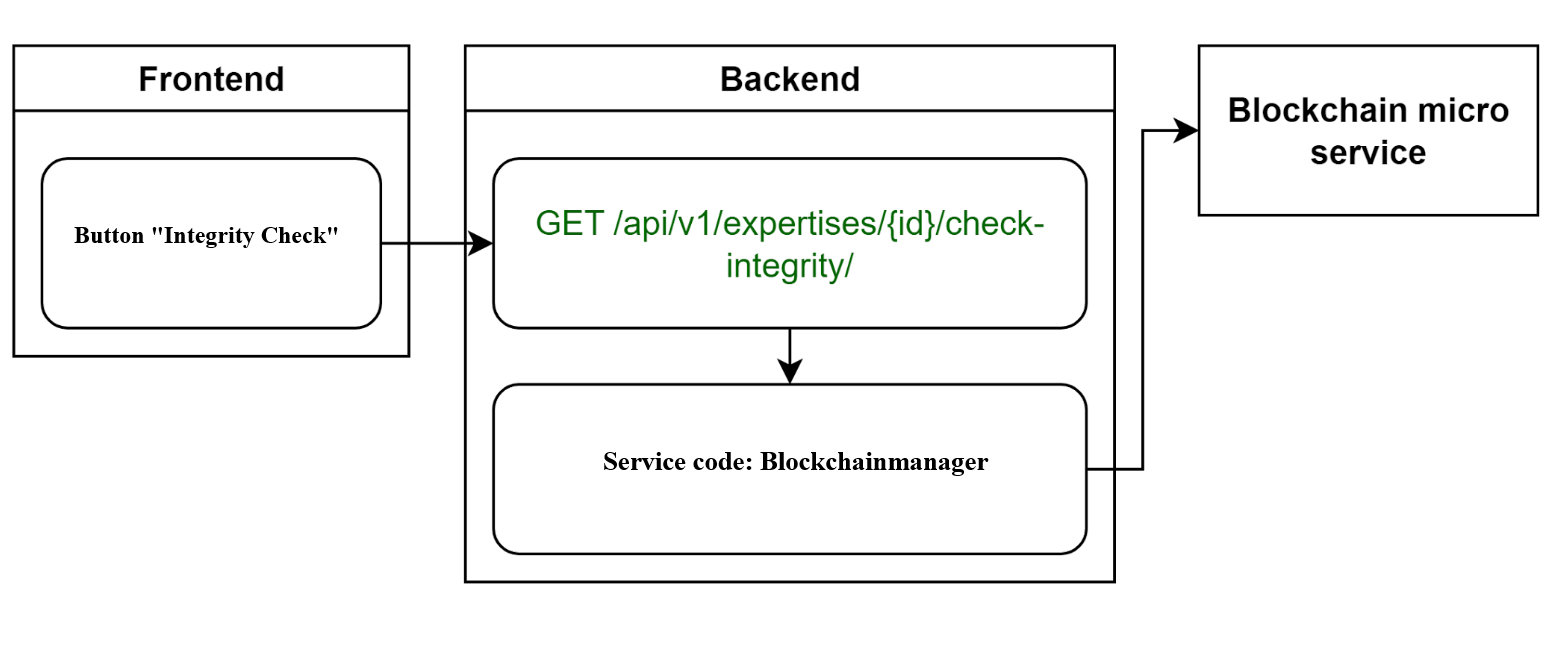}
\end{center}
\caption{ How Blockchain Microservice Integrated to our main system.}\label{fig:endpoint}
\end{figure}

The API endpoints provided for saving, retrieving, and checking the status of data play a pivotal role in enhancing the functionality and utility of our Blockchain project. By leveraging these endpoints, users can securely store sensitive information on the Blockchain, ensuring data integrity and confidentiality through encryption mechanisms. This capability not only enables transparent and auditable data management but also mitigates the risks associated with centralized storage systems, such as single points of failure and data tampering vulnerabilities. Additionally, the ability to retrieve and verify the status of transactions empowers users with real-time insights into the state of their data on the Blockchain, fostering trust and accountability in the system. Overall, these API endpoints contribute to the robustness, security, and usability of our Blockchain project, facilitating its adoption and utilization across diverse use cases and industries.
The main EPEC system uses a REST API to interact with our Blockchain microservice part, as shown in \ref{fig:endpoint}.

\subsection{Data Security and Encryption Approach}
Our implementation employs a combination of Gzip compression and Fernet symmetric encryption to ensure data security. This approach was selected for its simplicity, ease of implementation, and cost-effectiveness, aligning with our goal of creating a lightweight and accessible solution for educational institutions.

The encryption process begins with the plaintext message \( M \). This message is first compressed using Gzip, resulting in a compressed message \( M_c \):
\[ M_c = \text{Gzip}(M) \]
The compressed message \( M_c \) is then encrypted using Fernet with key \( k \), producing the final ciphertext \( C \):
\[ C = \text{Fernet}_k(M_c) \]
For decryption, we reverse these steps. First, the ciphertext \( C \) is decrypted using Fernet:
\[ M_c = \text{Fernet}_k^{-1}(C) \] Then, the compressed message \( M_c \) is decompressed using Gzip to recover the original message \( M \):
\[ M = \text{Gzip}^{-1}(M_c) \]

This method provides a good balance between security and performance for our current needs. However, we acknowledge that more advanced encryption techniques are available for scenarios requiring higher levels of security or privacy. For applications with more stringent privacy requirements, techniques such as Homomorphic Encryption (HE), Secure Multi-Party Computation (MPC), or Zero-Knowledge Proofs (ZKP) could be considered. Homomorphic Encryption, for instance, allows computations to be performed on encrypted data without decrypting it, offering enhanced privacy but at the cost of increased computational complexity. Our choice of Gzip compression and Fernet encryption was driven by several factors. The simplicity and robust documentation of these methods facilitate rapid development and deployment, crucial for our prototype implementation. The combination of compression and symmetric encryption provides a good balance between security and computational efficiency, which is particularly important in a Blockchain context where every operation has an associated gas cost. Moreover, for the current scope of EPEC, where the primary goal is to ensure the integrity of educational content rather than to process highly sensitive personal data, this level of encryption provides adequate security.

\subsection{Privacy and Security in Blockchain}
Before deploying a smart contract and utilizing a wallet to interface with the main project, it is standard practice to conduct testing and auditing procedures on the smart contract code. Employing a new wallet or implementing stringent privacy measures is recommended to mitigate potential privacy issues associated with Ethereum Virtual Machine (EVM) wallets. Recent studies have highlighted various privacy concerns in permissionless Blockchain environments, including the presence of malicious MVP (Minimum Viable Product) bots, vulnerabilities within smart contracts, and risks related to Remote Procedure Call (RPC) interfaces, \cite{bayan2023exploring}. These findings underscore the critical importance of implementing robust security protocols and maintaining ongoing vigilance to address emerging threats within the Blockchain ecosystem.

\section{Results}
\subsection{Performance Analysis}
\begin{figure}[h!]
\begin{center}
\includegraphics[width=16cm]{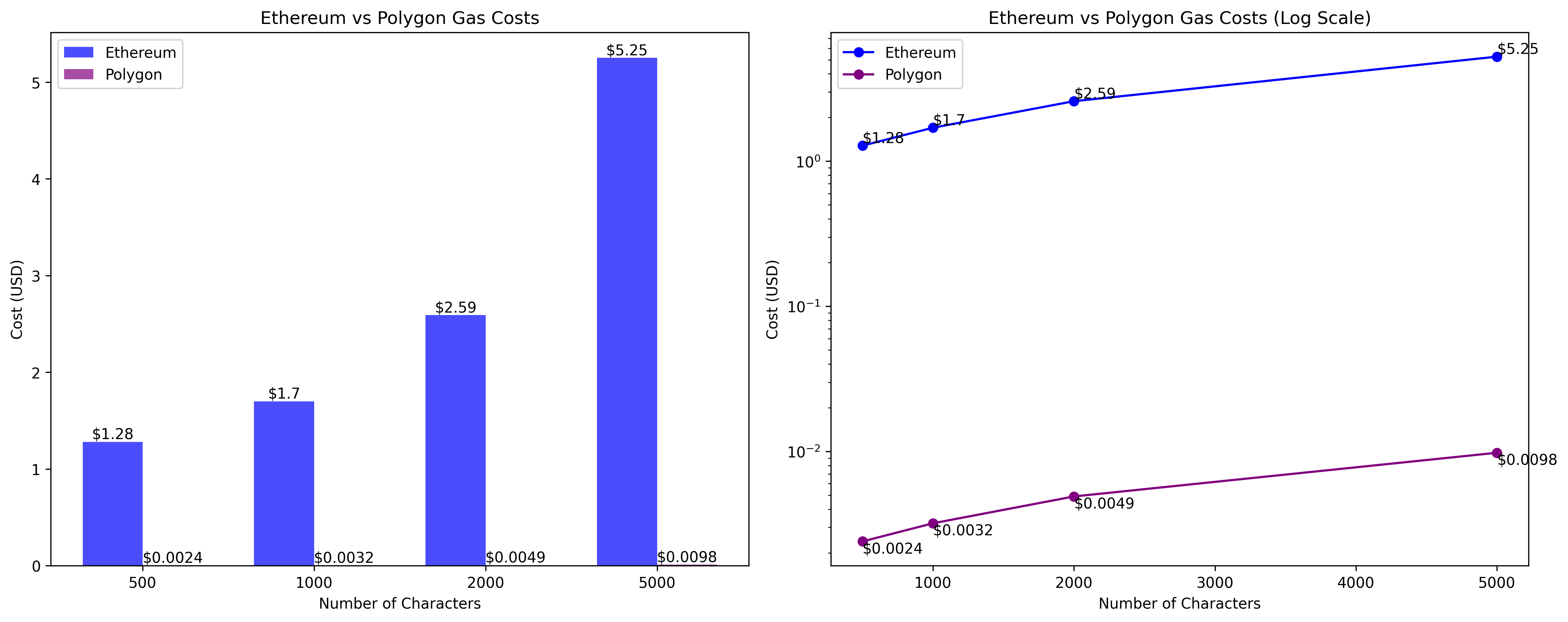}
\end{center}
\caption{ Comparison of gas costs between Ethereum and Polygon for different transaction sizes. (a) Bar chart comparison. (b) Logarithmic scale comparison.}\label{fig:2}
\label{fig:gas-cost-comparison}
\end{figure}

We calculated the gas costs for storing encrypted text of 500, 1000, 2000, and 5000 characters, using average gas prices and cryptocurrency values as of August 19, 2024. The calculations assumed a basic \texttt{store(bytes)} function in a smart contract with a base cost of 20,000 gas plus 16 gas per byte of data. Figure~\ref{fig:gas-cost-comparison} presents a visual comparison of estimated gas fees for Ethereum and Polygon. The figure consists of two charts: (a) a bar chart directly comparing costs, and (b) a logarithmic scale plot emphasizing the orders of magnitude difference between the two networks.
As evident from Figure~\ref{fig:gas-cost-comparison}, the cost difference between Ethereum and Polygon is substantial across all data sizes tested. For instance, storing 1000 characters of encrypted text costs approximately \$1.70 on Ethereum, compared to just \$0.0032 on Polygon --- a reduction of over 99.8\%.
Across all tested data sizes, smart contract on Polygon consistently achieved cost reductions exceeding 98\%, highlighting a key advantage of Layer 2 solutions in dramatically reducing transaction costs.
The scalability benefits of Polygon become increasingly apparent as data size grows. For instance, while storing 5000 characters on Ethereum costs \$5.25, the same operation on Polygon remains under \$0.01. This scalability is particularly crucial for applications like EPEC, which may require storage of large data volumes or frequent transactions. The logarithmic scale plot in Figure~\ref{fig:gas-cost-comparison}(b) illustrates that these relative cost savings remain consistent across different data sizes, suggesting reliable and predictable gas savings across various usage scenarios. In the context of EPEC, storing 1000 integrity verification records would cost approximately \$1,700 on Ethereum, but only \$3.20 on Polygon. Such cost-effectiveness makes Blockchain-based integrity verification more viable for large-scale implementations and potentially more accessible to a wider range of applications and users. This increased accessibility could drive greater adoption of Blockchain technology in educational content management systems and similar fields.

Why do Layer 2 solutions like Polygon achieve these significant gas savings through several mechanisms?  Most transaction processing occurs off the main Ethereum chain, reducing the computational load on the Ethereum network. Additionally, multiple transactions can be batched together and settled on the Ethereum mainnet as a single transaction, distributing the gas cost across many operations. These factors combine to dramatically reduce the gas fees required for each transaction, as clearly demonstrated in our analysis. Our gas savings analysis thus provides strong evidence for the cost-effectiveness of using Layer 2 solutions like Polygon for Blockchain-based data integrity verification. The substantial reduction in transaction costs not only makes our EPEC implementation more economically viable but also opens up possibilities for more frequent and comprehensive use of Blockchain technology in ensuring the integrity of educational content.

\subsection{Smart Contract Address and Interacting Wallet Security Considerations}
The transition from Web2 to Web3 technologies introduces novel security challenges, particularly in the context of permissionless Blockchains. Unlike traditional web applications, Blockchain platforms require users to manage their own \textbf{private keys} for interactions, thereby assuming full responsibility for security. This paradigm shift presents significant usability barriers, especially for users unfamiliar with Blockchain technologies.

We addressed these challenges by storing the smart contract address and the private key of the interacting wallet in a `.env` file during development. This approach simplifies user interaction but necessitates a robust key management system and centralized trust in the developers with access to these credentials. The decision to store sensitive credentials in a `.env` file was driven by several factors. Primarily, the majority of EPEC users are unfamiliar with Blockchain wallet management, including hardware wallets or browser extensions like MetaMask. This approach lowers the barrier to entry by eliminating the need for users to manage their private keys. Furthermore, our smart contract does not involve external calls or asset manipulations, mitigating some potential risks. Additionally, the system is intended for use by KYC-verified (Know Your Customer) workers and developers, providing an additional layer of accountability.

    A critical security concern in this implementation is the potential for a \textbf{"contract address spoofing"} attack, where a developer could alter the `.env` file to point to a duplicate smart contract. To mitigate this risk, we employ KYC-verified developers with established reputations and track records. Regular and thorough code reviews are conducted to detect any unauthorized changes, and strict version control practices are maintained to track and audit any modifications to critical files. While these measures provide a baseline of security, we acknowledge that more robust solutions could be implemented in future iterations of the system. Potential enhancements include embedding contract addresses directly in the application code, making unauthorized changes more difficult and detectable. Utilizing \textbf{Hardware Security Modules (HSMs)} to store contract addresses and sensitive data could significantly increase the difficulty of tampering. Additionally, deploying the application within \textbf{Trusted Execution Environments (TEEs)}, such as Intel SGX, could protect sensitive data from tampering, even in cases of system access.

\section{Discussion}

The integration of Blockchain technology, specifically the Polygon network, with EPEC demonstrates significant potential for enhancing data integrity verification in educational content management. Our results reveal several important implications for the field of educational technology. The immutability of Blockchain provides a robust defence against unauthorized alterations, while its public nature increases transparency in the verification process. This combination could significantly boost trust among stakeholders in the education ecosystem, including students, educators, and accreditation bodies. The ability to easily integrate with existing systems and selectively activate Blockchain features allows institutions to tailor the implementation to their specific needs and technical capabilities.

\subsection{Limitations and Future Work}
Our Blockchain-enhanced integrity verification system for educational content assessment, while promising, faces limitations in scalability with larger datasets and higher transaction volumes. Compliance with evolving educational and data protection regulations, as well as user interface optimization for non-technical users, remain ongoing challenges. Future work will focus on expanding platform capabilities and addressing these limitations. We plan to implement multi-chain support, allowing users to select preferred networks like Arbitrum, thus enhancing data immutability and flexibility. Our wallet integration strategy will evolve to support external interactions, such as MetaMask or hardware wallets, improving user control and security.

The development of sophisticated smart contracts, inspired by Decentralized Autonomous Organizations, will offer enhanced management of educational data with nuanced access control and governance structures. We will explore AI integration, particularly Large Language Models, for smart contract auditing to bolster security and reliability. To improve user engagement, we aim to develop a comprehensive dashboard within EPEC for visualizing Blockchain-related data. Integration of the InterPlanetary File System (IPFS) for decentralized storage of larger datasets is also planned, complementing our integrity verification system. As privacy concerns evolve, we will investigate advanced privacy-preserving techniques like zero-knowledge proofs. Cross-chain interoperability solutions will be explored to expand the system's reach and flexibility. Future research will examine Blockchain's role in the context of generative AI and frequent content updates in educational platforms. We will address the ethical implications of using Blockchain for immutable records and automated evaluations in education, ensuring alignment with pedagogical best practices and privacy concerns. These efforts aim to continually enhance our platform's security, efficiency, and user-friendliness, adapting to the evolving needs of educational institutions in the digital age.

\section{Conclusion}

This paper presents a novel approach to enhancing the integrity of educational content assessment through the integration of Blockchain technology with the Electronic Platform for Expertise of Content (EPEC). By leveraging the Polygon network, we have demonstrated a cost-effective and scalable solution for secure data storage and verification in educational contexts. Our implementation showcases significant advantages in terms of gas savings and transaction costs compared to the Ethereum mainnet, with cost reductions exceeding 98\%. This economic efficiency makes large-scale Blockchain adoption in educational settings more feasible. The use of smart contracts and encryption techniques ensures data privacy and integrity, while our modular design allows for flexible integration with existing systems.

While our current implementation shows promise, several challenges remain. These include the need for further scalability testing, optimizing user experience for non-technical users, and ensuring compliance with evolving data protection regulations. Future work will focus on multi-chain support, advanced smart contract development, and the integration of AI-assisted auditing tools. Moreover, future research should concentrate on developing practical frameworks for implementing Blockchain in teacher performance evaluation, exploring how Blockchain can be integrated into existing systems while addressing the technical and ethical challenges associated with its use.

In conclusion, this research demonstrates the potential of Blockchain technology to transform educational content management by enhancing transparency, security, and trust. As the field evolves, we anticipate that Blockchain-based solutions will play an increasingly important role in shaping the future of digital education, contributing to more efficient, fair, and secure assessment processes, particularly in the context of teacher evaluation and professional development.

\section*{Conflict of Interest Statement}

The authors declare that the research was conducted in the absence of any commercial or financial relationships that could be construed as a potential conflict of interest.

\section*{Author Contributions}
TB: Writing–original draft, Writing–review and editing, Software, Investigation, Methodology. RB: Methodology, Validation. AN: Methodology, Formal Analysis, Software. MG: Software, Visualization. AS: Software, Visualization. ZK: Writing–review and editing, Conceptualization, Investigation, Methodology, Supervision, Project Administration. 

\section*{Funding}
This research has been funded by the Committee of Science of the Ministry of Science and Higher Education of the Republic of Kazakhstan (Grant No. BR21882260).

\bibliographystyle{Frontiers-Harvard}
\bibliography{frontiers}


\end{document}